\documentclass[amsmath,amssymb,prd,notitlepage,longbibliography,superscriptaddress,nofootinbib]{revtex4-2}    
\usepackage{graphicx} 
\usepackage{float}
\usepackage{textcomp}
\usepackage{amsfonts,amsmath} 
\usepackage{mathrsfs} 
\usepackage{setspace}
\usepackage[utf8]{inputenc} 
\usepackage{comment} 
\usepackage{xcolor}
\usepackage{mathtools}
\usepackage{hyperref}
\makeatletter
\def\@fpheader{\relax}
\makeatother

\DeclareMathAlphabet{\mathbbold}{U}{bbold}{m}{n} 
\newcommand{\be}{\begin{equation}} \newcommand{\ee}{\end{equation}}

\newcommand{\thalf}{{\tfrac{1}{2}}}

\DeclareMathOperator{\Tr}{Tr}

\begin{document}

\title{QNMs of scalar fields on small Reissner-Nordstr\"{o}m-AdS$\mathbf{_5}$ black holes}

\author{Juli\'{a}n Barrag\'{a}n Amado}
  \email{Jose.Julian.Barragan.Amado@USherbrooke.ca}
    \affiliation{Department of Mathematics, University of Sherbrooke,
    2500, boul. de l'Universit\'{e}, Sherbrooke, Quebec, Canada}
  \affiliation{Departamento de F\'{i}sica, Universidade Federal de Pernambuco,
    50670-901, Recife, Pernambuco, Brazil}
  \affiliation{Van Swinderen Institute for Particle Physics and
    Gravity, University of Groningen, Groningen, The Netherlands} 
\author{Bruno Carneiro da Cunha}
  \email{bruno.ccunha@ufpe.br}
  \affiliation{Departamento de F\'{i}sica, Universidade Federal de Pernambuco, 
    50670-901, Recife, Pernambuco, Brazil} 
\author{Elisabetta Pallante}
  \email{e.pallante@rug.nl}
  \affiliation{Van Swinderen Institute for Particle Physics and
    Gravity, University of Groningen, Groningen, The Netherlands} 
  \affiliation{Nikhef, Science Park, Amsterdam, The Netherlands}

\begin{abstract}
We study the quasinormal modes (QNMs) of a charged scalar field on a Reissner-Nordstr\"{o}m-anti-de Sitter (RN-AdS$_{5}$) black hole in the small radius limit by using the isomonodromic method.
We also derive the low-temperature expansion of the fundamental QNM frequency. Finally, we provide numerical evidence that instabilities appear in the small radius limit for large values of the charge of the scalar field.
\end{abstract}


\maketitle

\section{Introduction}
\label{sec:1}

The energy extraction from a rotating uncharged black hole can be realized by the scattering of a wave of frequency $\omega$ and axial quantum number $m$ that satisfies the superradiance condition $\omega < m\Omega$, where $\Omega$ is the angular velocity of the black hole. It turns out that the reflected wave is amplified with respect to the original incoming wave \cite{Brito:2015oca}. By imposing generic boundary conditions at the horizon and infinity, multiple amplifications and reflections of the scattered wave can occur and lead to superradiant instabilities.

An analogous superradiance condition for charged fields on charged black holes can be given by $\omega < e\Phi$, where $e$ is the charge of the field and $\Phi$ the electrostatic potential difference between the horizon and the spatial infinity \cite{PhysRevD.7.949,Denardo:1973pyo,Gibbons:1975kk}. Then, by imposing an asymptotic AdS spacetime or a mirrorlike boundary condition, Reissner-Nordstr\"{o}m black holes can develop such instabilities, as it has been shown in four dimensions \cite{PhysRevD.81.124020,Uchikata:2011zz} and $d \geq 4$ dimensions in the small outer horizon radius, $r_{+}$, limit \cite{Wang:2014eha}.

Black-hole superradiance is of relevance in numerous areas, from high-energy physics to astrophysics, and to fundamental questions of General Relativity, such as the stability of a black-hole solution and the existence of hairy black hole configurations. In \cite{Gubser:2008px} it was suggested that a Reissner-Nordstr\"{o}m-AdS$_{4}$ black hole coupled to a Higgs field Lagrangian can exhibit a spontaneous Abelian gauge symmetry breaking due to a scalar condensate near the horizon, thus implying that superradiant instabilities can occur for a certain range of values of the charges and the masses of the black hole and the scalar field. This observation may also suggest a holographic description of spontaneous symmetry breaking and superfluidity through the gauge-gravity duality -- for comprehensive reviews see \cite{Hartnoll:2008kx,Hartnoll:2009sz}. In five dimensions, the phase transition between the RN-AdS$_{5}$ black holes and hairy black hole solutions at the onset of superradiant instabilities has been explored in \cite{Basu:2010uz,Dias:2011tj}.

We study the occurrence of instabilities in RN-AdS$_{5}$ by computing the QNMs frequencies of a charged scalar field scattering on a RN-AdS$_5$ black hole. Due to the boundary conditions at the black hole horizon and at spatial infinity, the QNMs are generically complex frequencies, where the real part is related to the oscillation frequency and the imaginary part gives the inverse of the damping time. A superradiant mode thus corresponds to an eigenfrequency with a positive imaginary part. 

In this context, we apply the isomonodromic method to recast the associated boundary value problem into an initial value problem for the corresponding isomonodromic tau function. Since the radial second-order ordinary differential equation possesses four regular singular points, see \eqref{eq:Heun}, one should expect that the initial conditions are given in terms of the Painlev\'{e} VI (PVI) tau function \cite{Amado:2017kao,Barragan-Amado:2018pxh}. Interestingly, in recent years, the Painlev\'{e} transcendents have been employed to solve a variety of problems such as correlation functions in Ising and Ashkin-Teller model
(see \cite{Gamayun:2013auu} and references therein), conformal maps \cite{Anselmo20180080}, black-hole physics \cite{Novaes:2014lha,daCunha:2015fna} and the Rabi model in quantum optics \cite{daCunha:2015npa}. The Painlevé transcendents implement the Riemann-Hilbert map relating the accessory parameters of the differential equations to the monodromy properties of its solutions, and \cite{Gamayun:2013auu,Gavrylenko:2016zlf} showed that their general expansion is expressible in terms of $c=1$ conformal blocks. We should mention at this point that an alternative proposal directly using four-dimensional supersymmetric gauge theories in the
Nekrasov-Shatashvili phase, corresponding to semi-classical conformal blocks, was developed in \cite{Aminov:2020yma,Bershtein:2021uts}. Also worthy of note is the effort to relate quantities of physical interest to monodromy data in \cite{Bonelli:2021uvf}, where the authors considered greybody factors, QNMs and Love numbers for the four-dimensional Kerr background.

As argued in \cite{Barragan-Amado:2018pxh}, in the small $r_{+}$ limit the first correction to the (fundamental) QNMs frequencies receives contributions from all the intermediate channels of the $c=1$ Virasoro conformal blocks expansion. Nevertheless, these contributions can be resummed by using the generating function for the Catalan numbers, allowing us to obtain an asymptotic expression for the frequencies \cite{Barragan-Amado:2018pxh}. 

This manuscript is organized as follows. In Sec. \ref{sec:2} we study charged scalar perturbations of the RN-AdS$_{5}$ black hole, and recast the eigenvalue problem into a set of transcendental equations involving the PVI tau function. Sec. \ref{sec:3} is devoted to the analysis of the QNMs in the small $r_{+}$ limit. We discuss two different regimes for the temperature of the black hole: the high-temperature and the low-temperature limit for small black holes. Firstly, we show that at high temperature the QNMs are given by the zeros of the PVI tau function, and compute an asymptotic expression in the small $r_{+}$ limit for the $\ell=0$ modes in Sec. \ref{sec:3a}. Secondly, in Sec. \ref{sec:3b} we provide an asymptotic expression for the fundamental QNM frequency in the low-temperature limit by applying the confluence limit of the PVI tau function. Sec. \ref{sec:3c} compares the asymptotic analytic expressions and the numerical solutions. We close in Sec. \ref{sec:4} with some remarks on the implications of the obtained results. Finally, in Appendix \ref{appendixA} we review the Nekrasov expansion of the PVI tau function.

\section{Scalar Fields in Reissner-Nordstr\"{o}m-$\mathbf{AdS_5}$}
\label{sec:2}
The line element of the Reissner-Nordstr\"{o}m-AdS$_{5}$ black hole is
\begin{equation}
ds^{2} = -f(r)dt^{2} + \dfrac{1}{f(r)}dr^{2} + r^{2}d\Omega_{3}^{2},
\end{equation}
where $d\Omega_{3}^{2}$ is the metric of the unit three-sphere and the function $f(r)$ is given by
\begin{equation}\label{eq:Delta_r}
f(r) = 1 -\frac{M}{r^{2}} + \frac{Q^{2}}{r^{4}} + r^{2} =\frac{\Delta_{r}}{r^{4}} =  \frac{(r^{2}-r_{+}^{2})(r^{2}-r_{-}^{2})(r^{2}-r_{0}^{2})}{r^{4}},
\end{equation}
with the AdS radius $L=1$ and the parameters $M$, $Q$ are related to the black hole mass and charge, respectively. There is a correspondence between the roots of $\Delta_{r}$ and the Killing horizons of the black hole, whose radial positions will be parametrized by the largest real root $r_+$, corresponding to the outer horizon.  The roots corresponding to $r_{-}^{2}$ and $r_{0}^{2}$ can be written as
\begin{equation}\label{eq:roots}
r_{-}^{2} = \frac{1}{2}\left(-1-r_{+}^{2}+\sqrt{(1+r_{+}^{2})^{2}+\frac{4Q^{2}}{r_{+}^{2}}}\right), \qquad r_{0}^{2} = \frac{1}{2}\left(-1-r_{+}^{2}-\sqrt{(1+r_{+}^{2})^{2}+\frac{4Q^{2}}{r_{+}^{2}}}\right).
\end{equation}
The temperature at each horizon is
\begin{equation}
T_{k} = \frac{1}{4\pi}\frac{d}{dr}\left(\frac{\Delta_{r}}{r^{4}}\right)\biggr\vert_{r=r_{k}}=\frac{1}{2\pi}\frac{(r_{k}^{2}-r_{i}^{2})(r_{k}^{2}-r_{j}^{2})}{r^{3}_{k}}, \qquad i,j \neq k,
\end{equation}
and specifically, at the outer horizon
\begin{equation}\label{eq:Tplus}
T_{+}=\dfrac{1}{2\pi}\dfrac{(r_{+}^{2}-r_{-}^{2})(r_{+}^{2}-r_{0}^{2})}{r_{+}^{3}},
\end{equation}
which by means of \eqref{eq:roots} reads
\begin{equation}
T_{+}=\dfrac{1}{2\pi}\left[\frac{1}{r_{+}}-\frac{Q^{2}}{r^{5}_{+}}+2r_{+}\right].
\end{equation}
For black holes satisfying the cosmic censorship conjecture, we require that $Q \leq Q_{c}$, where the critical charge $Q_{c}$ is the maximal charge at extremality $(T_{+}=0)$
\begin{equation}\label{eq:Qc}
Q_{c} = r_{+}^{2}\sqrt{1+2r_{+}^{2}}.
\end{equation}
We then parametrize the accessible charge values as $Q = q\,Q_{c}$, where $q \leq 1$ is an extremality parameter for fixed $r_{+} > 0$, and in terms of $q$ the roots $r_{-}^{2}$ and $r_{0}^{2}$ in \eqref{eq:roots} read 
\begin{equation}\label{eq:rminus_and_r0}
r_{-}^{2}=\frac{1}{2}\left(-1-r_{+}^{2}+\sqrt{1+2(1+2q^{2})r_{+}^{2}+(1+8q^{2})r_{+}^{4}}\right), \quad r_{0}^{2}=\frac{1}{2}\left(-1-r_{+}^{2}-\sqrt{1+2(1+2q^{2})r_{+}^{2}+(1+8q^{2})r_{+}^{4}}\right).
\end{equation}
As we will see in section \ref{sec:3}, one can think of small black holes near the critical charge $(q \lesssim 1)$ both at low temperature and at high temperature.

The electromagnetic potential of the charged black hole reads
\begin{equation}
A_{\mu}dx^{\mu} = \left(-\dfrac{\sqrt{3}}{2}\dfrac{Q}{r^{2}}+C\right)dt,
\end{equation}
where the choice of the constant $C$ is determined by the gauge choice \cite{Horowitz:2010gk}. We assume $C=0$ for a vanishing potential at the spatial infinity, while it has also been shown in \cite{Uchikata:2011zz}, for the case of asymptotically AdS, that the role of $C$ is merely a shift in the real part of the quasinormal frequency $\omega$, resulting in $\mathrm{Re}(\omega) \rightarrow \mathrm{Re}(\omega) + e\,C$. 

We consider massive charged scalar field perturbations of the RN-AdS$_{5}$ black hole. The field equation is the Klein-Gordon equation
\begin{equation}\label{eq:KG}
\dfrac{1}{\sqrt{-g}}D_{\mu}\left(\sqrt{-g}g^{\mu\nu}D_{\nu}\right)\Phi - \mu^{2}\Phi=0,
\end{equation}
where $D_{\mu} = \nabla_{\mu} - i\,eA_{\mu}$, and $e$ and $\mu$ are the field charge and mass, respectively. The scalar field solution of \eqref{eq:KG} can be decomposed by the Ansatz
\begin{equation}
\Phi_{\omega,\ell}(t,\theta,\varphi,\psi,r) = e^{-i\omega t}Y_{\ell}^{m_{1},m_{2}}(\theta,\varphi,\psi)R_{\omega,\ell}(r),
\end{equation}
where $Y_{\ell}^{m_{1},m_{2}}(\theta,\varphi,\psi)$ are the scalar (spinless) spherical harmonics on the three-sphere, with eigenvalues determined by the equation
\begin{equation}\label{eq:spheroidal_harmonics}
\Delta Y_{\ell}^{m_{1},m_{2}}(\theta,\varphi,\psi) = - \ell(\ell+2) Y_{\ell}^{m_{1},m_{2}}(\theta,\varphi,\psi),
\end{equation}
with $\ell$ the angular momentum quantum number and $m_{1}$ and $m_{2}$ are integers. By means of \eqref{eq:spheroidal_harmonics}, equation \eqref{eq:KG} reduces to a second-order ordinary differential equation for the radial function $R_{\omega,\ell}(r)$ of the form 
\begin{equation}\label{eq:radialODE}
\left[\frac{1}{r}\frac{d}{dr}\left(\frac{\Delta_{r}}{r}\frac{d}{dr}\right)+\frac{r^{6}}{\Delta_{r}}\left(\omega-\frac{\sqrt{3}}{2}\frac{e\,Q}{r^{2}}\right)^{2}-\mu^{2}r^{2}-\ell(\ell+2)\right]R_{\omega,\ell}(r)=0.
\end{equation}
Equation \eqref{eq:radialODE} possesses four regular singular points, located at the roots of $\Delta_{r}$ in \eqref{eq:Delta_r} and infinity. The characteristic exponents $\beta_{k},\beta_\infty$ are determined by the asymptotic behavior of the solution in \eqref{eq:radialODE} near the singular points
\begin{equation}\label{eq:betas}
\beta_{k}=\pm\frac{1}{2}\theta_{k}, \qquad \beta_{\infty}=\frac{1}{2}(2\pm\theta_{\infty}), \qquad\qquad k=+,-,0,
\end{equation}  
where
\begin{equation}\label{eq:thetas}
\theta_{k} = \frac{i}{2\pi\,T_{k}}\left(\omega-\frac{\sqrt{3}}{2}\frac{e\,Q}{r_{k}^{2}}\right), \qquad \theta_{\infty}=2-\Delta, \qquad\qquad k=+,-,0,
\end{equation}
and $\theta_{+}$ is the variation of the entropy $\delta S$ of the black hole as it absorbs a quantum of frequency $\omega$ and electric charge $e$ at the outer horizon
\begin{equation}
\theta_{+} = \frac{i}{2\pi}\delta S.
\end{equation} 
When $\delta S < 0$, i.e. $\mathrm{Im}\,\theta_{+} < 0$, unstable modes can occur and the wave function grows exponentially with time, which means that the black hole is unstable. Furthermore, $\theta_{\infty}$ is related to $\Delta$ -- the conformal dimension of the CFT primary field associated to the AdS$_{5}$ scalar, i.e. $\Delta(\Delta-4)=\mu^{2}$, with $\mu$ the mass of the scalar field.

One can reduce \eqref{eq:radialODE} to the canonical Heun form by performing the following change of variables:
\begin{subequations}
\begin{equation}\label{eq:Moebius}
z=\frac{r^{2}-r^{2}_{-}}{r^{2}-r^{2}_{0}}, 
\end{equation}
\begin{equation}\label{eq:s-homotopic}
R_{\omega,\ell}(z) = z^{-\theta_{-}/2}(z-z_{0})^{-\theta_{+}/2}(z-1)^{\Delta/2}F(z),
\end{equation}
\end{subequations}
that leads to an equation for $F(z)$
\begin{equation}\label{eq:Heun}
\frac{d^{2}F}{dz^{2}}+\left[\frac{1-\theta_{-}}{z}+\frac{1-\theta_{+}}{z-z_{0}}+\frac{\Delta-1}{z-1}\right]\frac{dF}{dz}+\left(\frac{\kappa_{1}\kappa_{2}}{z(z-1)}-\frac{z_{0}(z_{0}-1)K_{0}}{z(z-z_{0})(z-1)}\right)F(z)=0,
\end{equation}
with
\begin{equation}
\kappa_{1} = \frac{1}{2}(\theta_{-}+\theta_{+}-\Delta-\theta_{0}), \qquad \kappa_{2}=\frac{1}{2}(\theta_{-}+\theta_{+}-\Delta+\theta_{0}),
\end{equation}
and the accessory parameters $\{z_{0},K_{0}\}$ are 
\begin{subequations}\label{eq:heun_accessory}
\begin{equation}\label{eq:z00}
z_{0}=\frac{r_{+}^{2}-r_{-}^{2}}{r_{+}^{2}-r_{0}^{2}},
\end{equation}
\begin{equation}\label{eq:K0}
\begin{split}
4z_{0}(z_{0}-1)K_{0}=&-\frac{\ell(\ell+2)+\Delta(\Delta-4)r_{-}^{2}-\omega^{2}}{(r_{+}^{2}-r_{0}^{2})}-(z_{0}-1)\left[(\theta_{-}+\theta_{+}-1)^{2}-\theta_{0}^{2}-1\right] \\
&-z_{0}\left[2(\theta_{+}-1)(1-\Delta)+(2-\Delta)^{2}-2\right].
\end{split}
\end{equation}
\end{subequations}
The QNMs are solutions of \eqref{eq:Heun} obeying specific boundary conditions, in the sense that there is only an incoming wave at the horizon $r_{+}$ $(z=z_{0})$ and regularity at spatial infinity $(z=1)$. For the radial equation in \eqref{eq:radialODE} with $\mu^2>0$, the conditions read as follows:
 \begin{equation}
R_{\omega,\ell}(r)\sim
\begin{cases}
(r-r_{+})^{-\thalf\theta_{+}},& r\rightarrow r_{+}, \\
r^{2-\Delta},& r\rightarrow \infty, 
\end{cases}
\label{eq:boundaryforr}
\end{equation}
and $F(z)$ in \eqref{eq:Heun} is a regular function at the boundaries. The conditions in \eqref{eq:boundaryforr} will enforce the quantization of the (not necessarily real) frequencies $\omega$, the QNMs frequencies.

Following the seminal works of the Kyoto school \cite{Jimbo1981,JIMBO1981407,Jimbo:1982}, one of the authors has described in \cite{daCunha:2015fna}, by means of the method of isomonodromic deformations, the Riemann-Hilbert map between invariants of the representation of the monodromy group of the four-punctured Riemann sphere and Fuchsian $2 \times 2$ linear systems with four regular singular points. The monodromy group action on the solutions of the Fuchsian system is represented by matrices $M_{i}$ describing the analytic continuation as one circles once around a singular point $z_i$ of the system. The map gives a procedure to compute the accessory parameters $\lbrace z_{0},K_{0} \rbrace$ of the Heun equation \eqref{eq:Heun} in terms of the monodromy data  $\lbrace  \vec{\theta}, \vec{\sigma} \rbrace = \lbrace\theta_{1},\theta_{2},\theta_{3},\theta_{4};\sigma_{12}, \sigma_{23}, \sigma_{13} \rbrace$, where the single monodromy parameters are defined as follows
\begin{equation}
2 \cos \pi \theta_{i} = \Tr M_{i}, \qquad k = 1,2,3,4,
\end{equation}
and the composite monodromy parameters $\sigma_{ij}$ are given by 
\begin{equation}
2\cos \pi \sigma_{ij} = \Tr M_{i}M_{j}, \qquad i,j = 1,2,3, 
\end{equation}
where $M_{i}M_{j}$ represents the analytic continuation around two singular points. Following \cite{JIMBO1981407}, only two of the three composite monodromy parameters $\sigma_{ij} = \lbrace \sigma_{12}, \sigma_{23}, \sigma_{13} \rbrace$ are independent, and we choose the pair $(\sigma_{12}, \sigma_{23})$ in the solution of the differential equation \eqref{eq:Heun}. The parameter $\sigma_{12}$ from now on will be denoted by $\sigma$.

As shown in \cite{daCunha:2015fna} the Riemann-Hilbert map between the accessory parameters of the Heun differential equation \eqref{eq:Heun} and the monodromy parameters can be expressed implicitly via the PVI tau function
\begin{subequations}\label{eq:initial_value}
\begin{equation}\label{eq:toda}
\tau(\vec{\theta};\sigma,s;z_{0}) = 0,
\end{equation} 
\begin{equation}\label{eq:accessory}
  \left.
\frac{\partial}{\partial t}\log\tau(\vec{\theta}^{-};\sigma-1,s;t)\right|_{t=z_0} -\frac{(\theta_{2}-1)\theta_{3}}{2(z_{0}-1)}-\frac{(\theta_{2}-1)\theta_{1}}{2z_{0}} = K_{0}.
\end{equation}
\end{subequations}
where $\vec{\theta}^{-}=\lbrace \theta_1,\theta_2-1,\theta_3,\theta_4+1\rbrace$ and $s$ is a function of the monodromy parameters defined by \eqref{eq:sigma1t}.

Repeating the analysis of \cite{Barragan-Amado:2018pxh}, in order to solve the first condition \eqref{eq:toda}, one can invert the series expansion of the PVI tau function \eqref{eq:asymp_tau}, for $t=z_0$ sufficiently close to zero, to write
\begin{equation}\label{eq:Ipsilon}
\chi(\sigma;z_0) = \kappa\,z_{0}^{\sigma},
\end{equation}
where $\kappa$ is given in terms of the monodromy data by \eqref{eq:kappa}. From the structure of the tau function expansion, we can see that $\chi$ is analytic in $z_0$, and its expansion can be computed recursively from \eqref{eq:asymp_tau}, yielding
\begin{align} \chi(\sigma;z_0)=\biggl[&\frac{((\theta_{2}+\sigma)^2-\theta_1^2)((\theta_3+\sigma)^2-\theta^{2}_{4})}{16\sigma^2(\sigma- 1)^2}z_0\biggr]\biggl(1 + (1-\sigma)\frac{(\theta_1^2-\theta^{2}_{2})(\theta_3^2-\theta^{2}_{4})+\sigma^2(\sigma- 2)^2}{2\sigma^2(\sigma- 2)^2}z_0+{\cal O}(z_0^2)\biggr).
\label{eq:kappaexpansion}
\end{align}
We can then determine the logarithmic derivative of the tau function in \eqref{eq:accessory} using the expansion \eqref{eq:asymp_tau} and write the expansion of the accessory parameter $K_0$ in terms of the monodromy data and $z_0$
\begin{equation}\label{eq:fourt0K0}
\begin{gathered}
4z_{0}K_{0}=\left(\sigma-1\right)^{2}-\left(\theta_{1}+\theta_{2}-1\right)^{2}+\bigg[2(\theta_{3}-1)(\theta_{2}-1)
+\frac{\left((\sigma -1)^2+\theta_{2}^2-\theta_1^2-1\right)\left((\sigma-1)^2+\theta_3^2-\theta_{4}^2-1\right)}{2\sigma(\sigma-2)}\bigg]z_{0}\\
+\bigg[\frac{13}{32}\sigma(\sigma-2)+2\left(\theta_{3}-1\right)\left(\theta_{2}-1\right)-\frac{1}{32}\left(5+14(\theta^2_{1}+\theta^{2}_{4})-18(\theta^{2}_{2}+\theta^{2}_{3})\right) +\frac{\left(\theta^{2}_{1}-\theta^{2}_{2}\right)^{2}\left(\theta^{2}_{3}-\theta^{2}_{4}\right)^{2}}{64}\left(\frac{1}{\sigma^{3}}-\frac{1}{(\sigma-2)^{3}}\right)\\
-\frac{\left((\theta^{2}_{1}-\theta^{2}_{2})(\theta^{2}_{3}-\theta^{2}_{4})+8\right)^{2}-2(\theta^{2}_{1}+\theta^{2}_{2})(\theta^{2}_{3}-\theta^{2}_{4})^{2}-2(\theta^{2}_{1}-\theta^{2}_{2})^{2}(\theta^{2}_{3}+\theta^{2}_{4})-64}{32\sigma(\sigma-2)}\\
+\frac{\left((\theta_{1}-1)^{2}-\theta^{2}_{2}\right)\left((\theta_{1}+1)^{2}-\theta^{2}_{2}\right)\left((\theta_{3}-1)^{2}-\theta^{2}_{4}\right)\left((\theta_{3}+1)^{2}-\theta^{2}_{4}\right)}{32(\sigma+1)(\sigma-3)}
\bigg]z^{2}_{0} +\mathcal{O}(z_{0}^{3}).
\end{gathered}
\end{equation}
Equations \eqref{eq:kappaexpansion} and \eqref{eq:fourt0K0} were derived under the assumption that $\sigma$ lies in the strip $0< \mathrm{Re}\,\sigma < 1$. Formulas for ${\rm Re}\,\sigma<0$ can be obtained by the replacement $\sigma \rightarrow -\sigma$. The expansion for $K_0$ is symmetric under the replacement $\sigma \rightarrow 2-\sigma$ and has the generic structure
\begin{equation}\label{eq:generickn}
4z_{0}K_{0} = k_{0} + k_{1}z_{0} + k_{2}z_{0}^{2} + \ldots + k_{n}z_{0}^{n} + \ldots,
\end{equation}
where $k_{n}$ is a rational function of the monodromy parameters, the numerator is a polynomial in the monodromy parameters $\theta_{i}$ and $\sigma$, and the denominator is a polynomial of $\sigma$ alone. The coefficient $k_n$ has single poles at integer $\sigma=3,\ldots,n+1$, as well as poles of order up to $2n-1$ at $\sigma=2$. The leading behavior of $k_n$ near $\sigma=2$ is
\begin{equation}\label{eq:kn}
k_{n} = (-1)^{n-1}\mathbf{C}_{n-1}\frac{(\theta_{2}^{2}-\theta_{1}^{2})^{n}(\theta_{3}^{2}-\theta_{4}^{2})^{n}}{4^{2n-1}(\sigma-2)^{2n-1}}+\ldots, \qquad n \geq 1,
\end{equation}
where $\mathbf{C}_{n}$ is the $n$-th Catalan number. A structure analogous to \eqref{eq:generickn} exists for $\chi(\sigma;z_{0})$ in \eqref{eq:kappaexpansion}, and the reader is referred to \cite{Barragan-Amado:2018pxh} for further details. Recently, similar analytic expansions involving irregular conformal blocks have been analyzed in \cite{CarneirodaCunha:2019tia,daCunha:2021jkm}.

For our particular problem of the radial system, the single monodromy parameters $\theta_i$ can be read directly from \eqref{eq:Heun} as follows
\begin{equation}\label{eq:localmono}
\theta_{1} = \theta_{-}, \quad \theta_{2} = \theta_{+}, \quad \theta_{3} = 2-\Delta, \quad \theta_{4} = \theta_{0},
\end{equation}
and then the conditions \eqref{eq:toda} and \eqref{eq:accessory} define for us the monodromy parameters $\sigma$ and $\kappa$ (or, rather, $s$) as functions of the modulus $z_0$ and the accessory parameter $K_0$. As it was shown in \cite{Novaes:2014lha}, physically relevant quantities such as scattering coefficients can be expressed in terms of $\sigma$ and $s$.

For the QNM problem at hand, we have to enforce the boundary conditions in \eqref{eq:boundaryforr}. As shown in \cite{Barragan-Amado:2018pxh}, these can be encoded with the requirement that the composite monodromy matrix around the two singular points $r_{+}$ and $\infty$ is triangular. In turn, this implies that the composite monodromy parameter $\sigma_{23}$ satisfies
\begin{equation}\label{eq:radial_cond}
\sigma_{23} = \theta_{2}+\theta_{3}+2n, \qquad n \in \mathbb{Z},
 \end{equation}
which will be referred to as the quantization condition for the radial function. With this restriction, we can solve for the $s$ parameter in \eqref{eq:sigma1t} to yield 
\begin{equation}\label{eq:ese}
s = \frac{\sin\tfrac{\pi}{2}(\theta_{2}+\theta_{1}+\sigma)\sin\tfrac{\pi}{2}(\theta_{2}-\theta_{1}+\sigma)\sin\tfrac{\pi}{2}(\theta_{3}+\theta_{4}+\sigma)\sin\tfrac{\pi}{2}(\theta_{3}-\theta_{4}+\sigma)}{\sin\tfrac{\pi}{2}(\theta_{2}+\theta_{1}-\sigma)\sin\tfrac{\pi}{2}(\theta_{2}-\theta_{1}-\sigma)\sin\tfrac{\pi}{2}(\theta_{3}+\theta_{4}-\sigma)\sin\tfrac{\pi}{2}(\theta_{3}-\theta_{4}-\sigma)},
\end{equation}
which, given the relation between $\kappa$ and $s$ in \eqref{eq:kappa}, can then be substituted in \eqref{eq:Ipsilon} to give
\begin{equation}\label{eq:painleve}
\frac{\Gamma^{2}(1-\sigma)}{\Gamma^{2}(1+\sigma)}\frac{\Gamma(1+\tfrac{1}{2}(\theta_{2}+\theta_{1}+\sigma))}{\Gamma(1+\tfrac{1}{2}(\theta_{2}+\theta_{1}-\sigma))}\frac{\Gamma(1+\tfrac{1}{2}(\theta_{2}-\theta_{1}+\sigma))}{\Gamma(1+\tfrac{1}{2}(\theta_{2}-\theta_{1}-\sigma))}\frac{\Gamma(1+\tfrac{1}{2}(\theta_{3}+\theta_{4}+\sigma))}{\Gamma(1+\tfrac{1}{2}(\theta_{3}+\theta_{4}-\sigma))}\frac{\Gamma(1+\tfrac{1}{2}(\theta_{3}-\theta_{4}+\sigma))}{\Gamma(1+\tfrac{1}{2}(\theta_{3}-\theta_{4}-\sigma))}s\,z_{0}^{\sigma} = \chi(\sigma;z_{0}).
\end{equation}
Equations \eqref{eq:fourt0K0} and \eqref{eq:painleve} can be interpreted as an overdetermined system for $\sigma$ in terms of the parameters of the differential equation, namely $z_0$ in \eqref{eq:z00}, $K_0$ in \eqref{eq:K0} and $\vec{\theta}$ in \eqref{eq:localmono}, which will only admit solutions for specific (discrete) values of $\omega$. As we will see in the following, these equations can be solved perturbatively in the small $r_+$ limit to provide us with the QNMs frequencies.

Similar methods have been employed to study scalar and vector perturbations in Kerr-AdS$_{5}$ backgrounds \cite{Barragan-Amado:2018pxh,Amado:2020zsr}, and more recently in the near-extremal limit \cite{BarraganAmado:2021uyw}. In four dimensions, rotating black holes in asymptotically flat and de-Sitter spacetimes were analyzed in \cite{Novaes:2018fry,CarneirodaCunha:2019tia,daCunha:2021jkm}.

\section{Small RN-AdS$_{5}$ black holes via isomonodromy}
\label{sec:3}

We proceed to study the small $r_{+}$ limit of RN-AdS$_{5}$ black holes. By means of \eqref{eq:rminus_and_r0}, the accessory parameter $z_{0}$ in \eqref{eq:z0} can be rewritten in terms of $r_{+}$ and $q$ only
\begin{equation}\label{eq:z0}
z_{0} = \frac{r_{+}^{2}-r_{-}^{2}}{r_{+}^{2}-r_{0}^{2}} = \frac{1+3r_{+}^{2}-\sqrt{1+2(1+2q^{2})r_{+}^{2}+(1+8q^{2})r_{+}^{4}}}{1+3r_{+}^{2}+\sqrt{1+2(1+2q^{2})r_{+}^{2}+(1+8q^{2})r_{+}^{4}}}.
\end{equation}
For small $r_{+}$ and arbitrary $0 \leq q \leq 1$
\begin{equation}\label{eq:z0_small}
z_{0} = (1-q^{2})r_{+}^{2} + \mathcal{O}(r_{+}^{4}),
\end{equation}
so that $r_{+} \rightarrow 0$ implies $z_{0} \rightarrow 0$ in \eqref{eq:z0_small}. Hence, the analytic expansions for $\chi(\sigma;z_{0})$ in \eqref{eq:kappaexpansion} and the accessory parameter $K_{0}$ in \eqref{eq:fourt0K0} for small $z_0$ apply in the small $r_{+}$ limit. 

By means of \eqref{eq:rminus_and_r0} and $q = Q/Q_{c}$, the temperature \eqref{eq:Tplus} reduces to
\begin{equation}\label{eq:twopitemp}
  2\pi T_{+} = \frac{1-q^{2}}{r_{+}} + 2(1-q^{2})r_{+}=
  \frac{\epsilon(2-\epsilon)}{r_+}+2\epsilon(2-\epsilon)r_+,
\end{equation}
where in the second equality we introduced the extremality parameter $\epsilon=1-q$. For
small black holes $r_+\ll 1$ and $\epsilon$ fixed, the terms of order $1/r_{+}$ in the last equality of \eqref{eq:twopitemp} dominate as $r_{+} \rightarrow 0$.

Equation \eqref{eq:twopitemp} also suggests that we can identify two different regimes, depending on the behavior of $\epsilon$:
\begin{enumerate}
\item The first regime corresponds to $0 < \epsilon < 1$ fixed and small black hole, $r_{+} \ll 1$, hence high temperature $T_{+} \sim 1/r_{+}$. Specifically, we should set $r_{+} \ll \epsilon < 1$ in our expansions. This regime is investigated in Sec. \ref{sec:3a}. 
\item The second regime describes the scaling limit for $\epsilon \ll r_{+} \ll 1$, as both $\epsilon$ and $r_{+}$ vary. This is the low temperature regime and it is explored in Sec. \ref{sec:3b}.
\end{enumerate}

\subsection{QNMs frequencies at high temperature}
\label{sec:3a}
The single monodromy parameters $\theta_{\pm}$ and $\theta_{0}$ can be expanded for small $r_{+}$ as follows
\begin{align}\label{eq:smallr2_monodromies}
\theta_{-} &= -i\frac{q^{2}\left(2q\omega-\sqrt{3}e\right)}{2(1-q^{2})}r_{+} + \mathcal{O}(r_{+}^{3}), \nonumber \\
\theta_{+} &= i\frac{\left(2\omega-\sqrt{3}eq\right)}{2(1-q^{2})}r_{+} + \mathcal{O}(r_{+}^{3}), \\
\theta_{0} &= \omega + \frac{1}{2}\left(\sqrt{3}eq-3(1+q^{2})\omega\right)r_{+}^{2} + \mathcal{O}(r_{+}^{4}), \nonumber
\end{align}
with
\begin{equation}\label{eq:varthetas}
\theta_{-} = -i\vartheta_{-}r_{+}, \qquad \theta_{+} = i\vartheta_{+}r_{+},
\end{equation}
and the coefficients $\vartheta_{\pm}$ are finite for $q < 1$ fixed, such that $\theta_{\pm}$ approach zero as $r_{+} \rightarrow 0$.
The expansion \eqref{eq:fourt0K0} for $z_{0}K_{0}$ allows us to compute the composite monodromy parameter $(\sigma=\sigma_{\ell})$ by inverting the series in \eqref{eq:fourt0K0}. For $\ell \geq 2$ the result is
\begin{align}\label{eq:sigma0t}
\sigma_{\ell} &= \sum_{n=0}^{\infty} c_{n}r_{+}^{n} \nonumber \\
&= \ell+2-\frac{((1+q^{2})(3\omega^{2}+3\ell(\ell+2)-\Delta(\Delta-4))-2\sqrt{3}eq\omega)}{4(\ell+1)}r_{+}^{2} + \mathcal{O}(r_{+}^{4}), \qquad \ell \geq 2.
\end{align}
For $\ell=0$ and $\ell=1$, the pole structure of $k_n$ described in \eqref{eq:generickn} renders the derivation of the coefficients $c_n$ more complicated, and we will deal separately with the $\ell=0$ case below. At any rate, $\sigma_\ell$ always admits the generic structure
\begin{equation}\label{eq:sigmaell}
\sigma_{\ell} = \ell+2-\nu_{\ell}r_{+}^{2} + \mathcal{O}(r_{+}^{4}).
\end{equation} 
By substituting \eqref{eq:sigma0t} in \eqref{eq:kappaexpansion}, one obtains an expansion for small $r_{+}$ for $\chi(\sigma_{\ell};z_{0})$
\begin{equation}\label{eq:chiell}
\chi_{\ell} \equiv \chi(\sigma_{\ell};z_{0}) =-(1-q^{2})\frac{\omega^{2}-(\Delta-\ell-4)^{2}}{16(\ell+1)^{2}}\left(1+i\frac{2\vartheta_{+}}{(\ell+2)}r_{+}\right)r_{+}^{2}+\mathcal{O}(r_{+}^{4}), \qquad \ell \geq 2,
\end{equation}
where $\ell \geq 2$ avoids the poles in \eqref{eq:kappaexpansion}. 

We now focus on the $\ell = 0$ case, and defer the analysis of higher $\ell$ modes to future work.
The $\ell=0$ case is peculiar, since the leading behavior of $\sigma_{0} - 2$ in \eqref{eq:sigmaell} is of order $r_{+}^{2}$ and the scaling limit of $\theta_{\pm}$ in \eqref{eq:varthetas} suggests that the accessory parameter $K_{0}$ receives contributions from all orders in $z_{0}$ in the small $r_{+}$ limit. The left hand side of \eqref{eq:fourt0K0} is given by \eqref{eq:K0} for $\ell=0$, and for small $r_{+}$ we have 
\begin{equation}
4z_{0}K_{0} = -2i(\vartheta_{-}-\vartheta_{+})r_{+} + \left((\vartheta_{-}-\vartheta_{+})^{2} + \Delta^{2} - 2(1+q^{2})\Delta + \omega(\sqrt{3}eq-3(1+q^{2})\omega) + (1+2q^{2})\omega^{2}\right)r_{+}^{2} + \mathcal{O}(r_{+}^{3}),
\end{equation}
whereas the right hand side of \eqref{eq:fourt0K0} for $\ell=0$ is
\begin{align}
4z_{0}K_{0} &= -2i(\vartheta_{-}-\vartheta_{+})r_{+} + \biggl[(\vartheta_{-}-\vartheta_{+})^{2}-2\nu_{0}-\frac{1}{2}(1-q^{2})(\omega^{2}-\Delta^{2}) \nonumber \\
&+\frac{(1-q^{2})(\omega^{2}-(\Delta-2)^{2})(\vartheta_{-}^{2}-\vartheta_{+}^{2})}{4\nu_{0}}+\frac{(1-q^{2})^{2}(\omega^{2}-(\Delta-2)^{2})^{2}(\vartheta_{-}^{2}-\vartheta_{+}^{2})^{2}}{64\nu_{0}^{3}} \nonumber \\
&+\frac{(1-q^{2})^{3}(\omega^{2}-(\Delta-2)^{2})^{3}(\vartheta_{-}^{2}-\vartheta_{+}^{2})^{3}}{512\nu_{0}^{5}} + \frac{5(1-q^{2})^{4}(\omega^{2}-(\Delta-2)^{2})^{4}(\vartheta_{-}^{2}-\vartheta_{+}^{2})^{4}}{16384\nu_{0}^{7}} +\ldots\biggr]r_{+}^{2} + \mathcal{O}(r_{+}^{3}),
\end{align}
where $\nu_{0} = \nu_{\ell}(\ell=0)$ in \eqref{eq:sigmaell} and the dots are higher orders in $z_{0}$ that contribute to the $r_{+}^{2}$-term. It was shown in \cite{Barragan-Amado:2018pxh} that all the contributions from the conformal blocks expansion can be resummed using the generating function for the Catalan numbers to compute the first correction to $\sigma_{0}$ of order $r_{+}^{2}$. By defining 
\begin{equation}\label{eq:big_x}
X = \frac{(1-q^{2})(\omega^{2}-(\Delta-2)^{2})(\vartheta_{-}^{2}-\vartheta_{+}^{2})}{16\nu_{0}^{2}},
\end{equation}
and resumming all the terms in the series will lead to an equation for the coefficient $\nu_{0}$
\begin{equation}\label{eq:eqnnu0}
\nu_{0}\sqrt{1-4X} = \frac{1}{2}\omega(3(1+q^{2})\omega-\sqrt{3}eq)-\frac{1}{4}(1+q^{2})(3\omega^{2}-\Delta(\Delta-4)) + \mathcal{O}(r_{+}^{2}).
\end{equation}
Then, $\nu_{0}$ in \eqref{eq:eqnnu0} admits the following small $r_{+}$ expansion
\begin{equation}\label{eq:nu0}
\begin{split}
&\nu_{0} = \frac{1}{4}\biggl[(1+q^2)^{2}((3\omega^2-\Delta(\Delta-4))^{2}-4\omega^{2}(\omega^{2}-(\Delta-2)^{2}))+ 4q^{2}\omega^{2}(\omega^{2}-(\Delta-2)^{2}) \\
&\qquad\quad +3e^{2}q^{2}(3\omega^{2}+(\Delta-2)^{2})-8\sqrt{3}(1+q^2)eq\omega(\omega^{2}+2)\biggr]^{\!1/2} + \mathcal{O}(r_{+}^{2}).
\end{split}
\end{equation}
Hence, by means of \eqref{eq:big_x} and resumming all the contributions of order $r_{+}^{2}$, we get for $\chi_{0}$ in \eqref{eq:chiell}
\begin{equation}\label{eq:chi_0}
\chi_{0} =-(1-q^{2})\frac{(\omega^{2}-(\Delta-4)^{2})}{64}r^{2}_{+}(1+i\vartheta_{+}r_{+})\left(1+\sqrt{1-\frac{(1-q^2)(\omega^{2}-(\Delta-2)^{2})(\vartheta_{-}^{2}-\vartheta_{+}^{2})}{4\nu_{0}^{2}}}\right)^{2}+\mathcal{O}(r_{+}^{4}),
\end{equation}
where
\begin{equation}
\vartheta_{-}=\frac{q^{2}\left(2q\omega-\sqrt{3}e\right)}{2(1-q^{2})}, \qquad \vartheta_{+}=\frac{\left(2\omega-\sqrt{3}eq\right)}{2(1-q^{2})}.
\end{equation}
The knowledge of the parameter $s$ in \eqref{eq:painleve} is now sufficient to determine the QNMs frequencies. By expanding for small $r_{+}$ and employing \eqref{eq:chi_0} in \eqref{eq:painleve}, we obtain
\begin{equation}\label{eq:ess0}
s = \Sigma_{0}\left(1+\frac{i2\nu_{0}\vartheta_{+}r_{+}}{(\vartheta_{+}^{2}-\vartheta_{-}^{2})}\right) + \mathcal{O}(r_{+}^{2},r_{+}^{2}\log r_{+}),
\end{equation}
with
\begin{equation}
\Sigma_{0} = \nu_{0}^{2}\frac{\Gamma(\tfrac{1}{2}(2-\omega-\Delta))\Gamma(\tfrac{1}{2}(2+\omega-\Delta))}{\Gamma(\tfrac{1}{2}(6-\omega-\Delta))\Gamma(\tfrac{1}{2}(6+\omega-\Delta))}\frac{(\omega^{2}-(\Delta-4)^{2})}{4(1-q^2)(\vartheta_{+}^{2}-\vartheta_{-}^{2})}\left(1+\sqrt{1+\frac{(1-q^2)(\omega^{2}-(\Delta-2)^{2})(\vartheta_{+}^{2}-\vartheta_{-}^{2})}{4\nu_{0}^{2}}}\right)^{2}.
\end{equation}
Equation \eqref{eq:ess0} is a non-analytic expansion in $r_{+}$ due to terms $\mathcal{O}(r_{+}^{2}\log r_{+})$, which are originated by the term $z_{0}^{\sigma_{0}}$ in \eqref{eq:painleve}. However, the limit as $r_{+} \rightarrow 0$ is finite. 

We now parametrize the QNMs frequencies $\omega_{n,\ell}$ with $n \geq 0$ and $\ell=0$ in terms of a perturbation to the normal modes in empty AdS$_{5}$
\begin{equation}\label{eq:omega}
\omega_{n,0} = 2n + \Delta + \eta_{n,0}r_{+}^{2}, 
\end{equation}
under the assumption that $\eta_{n,0}$ has a finite limit as $r_{+} \rightarrow 0$. Then, by replacing \eqref{eq:omega} in \eqref{eq:smallr2_monodromies} for $\omega$, one obtains
\begin{equation}\label{eq:thetan}
\begin{split}
\theta_{-} &= -i\frac{q^{2}\left(2q(2n+\Delta)-\sqrt{3}e\right)}{2(1-q^{2})}r_{+} + \mathcal{O}(r_{+}^{3}), \\
\theta_{+} &= i\frac{\left(2(2n+\Delta)-\sqrt{3}eq\right)}{2(1-q^{2})}r_{+} + \mathcal{O}(r_{+}^{3}), \\
\theta_{0} &= 2n + \Delta + \beta_{n,0}r_{+}^{2} + \mathcal{O}(r_{+}^{4}),
\end{split}
\end{equation}
where $\beta_{n,0}$ encodes the correction $\eta_{n,0}$ to the QNMs as follows
\begin{equation}\label{eq:first_correction}
\beta_{n,0} = \eta_{n,0} + \frac{1}{2}\left(\sqrt{3}eq-3(1+q^2)(2n+\Delta)\right).
\end{equation} 
On the other hand, $\beta_{n,0}$ can also be computed by equating \eqref{eq:ese} to \eqref{eq:ess0}, where $\Sigma_{n,0} \equiv \Sigma_{0}(\omega = \omega_{n,0})$ and the radial monodromies are given by \eqref{eq:thetan}, and by expanding for small $r_{+}$ we have
\begin{equation}\label{eq:betan0}
\beta_{n,0} = \nu_{n,0}\frac{1+\Sigma_{n,0}}{1-\Sigma_{n,0}} + 4i\frac{\vartheta_{+}\nu_{n,0}^{2}}{(\vartheta_{-}^{2}-\vartheta_{+}^{2})}\frac{\Sigma_{n,0}}{(\Sigma_{n,0}-1)^{2}}r_{+} + \mathcal{O}(r_{+}^{2},r_{+}^{2}\log r_{+}),
\end{equation}
with $\nu_{n,0} \equiv \nu_{0}(\omega = \omega_{n,0})$ in \eqref{eq:nu0}. By means of \eqref{eq:betan0}, \eqref{eq:first_correction} yields for $\eta_{n,0}$
\begin{equation}\label{eq:etan0}
\eta_{n,0} = \frac{1}{2}\left(3(1+q^2)(2n+\Delta)-\sqrt{3}eq-Z_{n,0}^{1/2}-i(n+1)(\Delta+n-1)(2(2n+\Delta)-\sqrt{3}eq)r_{+}\right) + \mathcal{O}(r_{+}^{2}, r_{+}^{2}\log r_{+}),
\end{equation}
where
\begin{equation}
Z_{n,0} = \left(\frac{1}{2}(1+q^2)(3(2n+\Delta)^{2}-\Delta(\Delta-4))-\sqrt{3}eq(2n+\Delta)\right)^{2}.
\end{equation}
Finally, by employing \eqref{eq:etan0}, \eqref{eq:omega} admits the small $r_{+}$ expansion
\begin{equation}\label{eq:omegan0}
\begin{split}
&\omega_{n,0} = 2n + \Delta -\frac{1}{2}\left(Z_{n,0}^{1/2}-3(1+q^2)(2n+\Delta)+\sqrt{3}eq\right)r_{+}^{2}-i\frac{1}{2}(n+1)(\Delta+n-1)(2(2n+\Delta)-\sqrt{3}eq)r_{+}^{3} \\
&\qquad\quad + \mathcal{O}(r_{+}^{4},r_{+}^{4}\log r_{+}).
\end{split}
\end{equation}
In fact, equation \eqref{eq:omegan0} is different from the analytical result in \cite{Wang:2014eha} due to the correction to the real part of $\omega_{n,0}$ of order $r_{+}^{2}$. The asymptotic expansion \eqref{eq:omegan0} for $n=0$ reproduces the numerical results for the real part of $\omega_{0,0}$ of Table $\mathrm{I}$ in \cite{Wang:2014eha}. In addition, the imaginary part of $\omega_{n,0}$ for $n=0$ in \eqref{eq:omegan0} for specific values of $e$, $\Delta$, $q$ and $r_{+}$ falls in between the numerical and analytical values given in Tables $\mathrm{II,IV,V}$  for the five-dimensional case in \cite{Wang:2014eha}. 
Importantly, for $q > \tfrac{2(2n+\Delta)}{\sqrt{3}e}$, the imaginary part in \eqref{eq:omegan0} becomes positive, and the superradiance condition is satisfied. 

\begin{figure}
\centering
\includegraphics[width=0.6\textwidth]{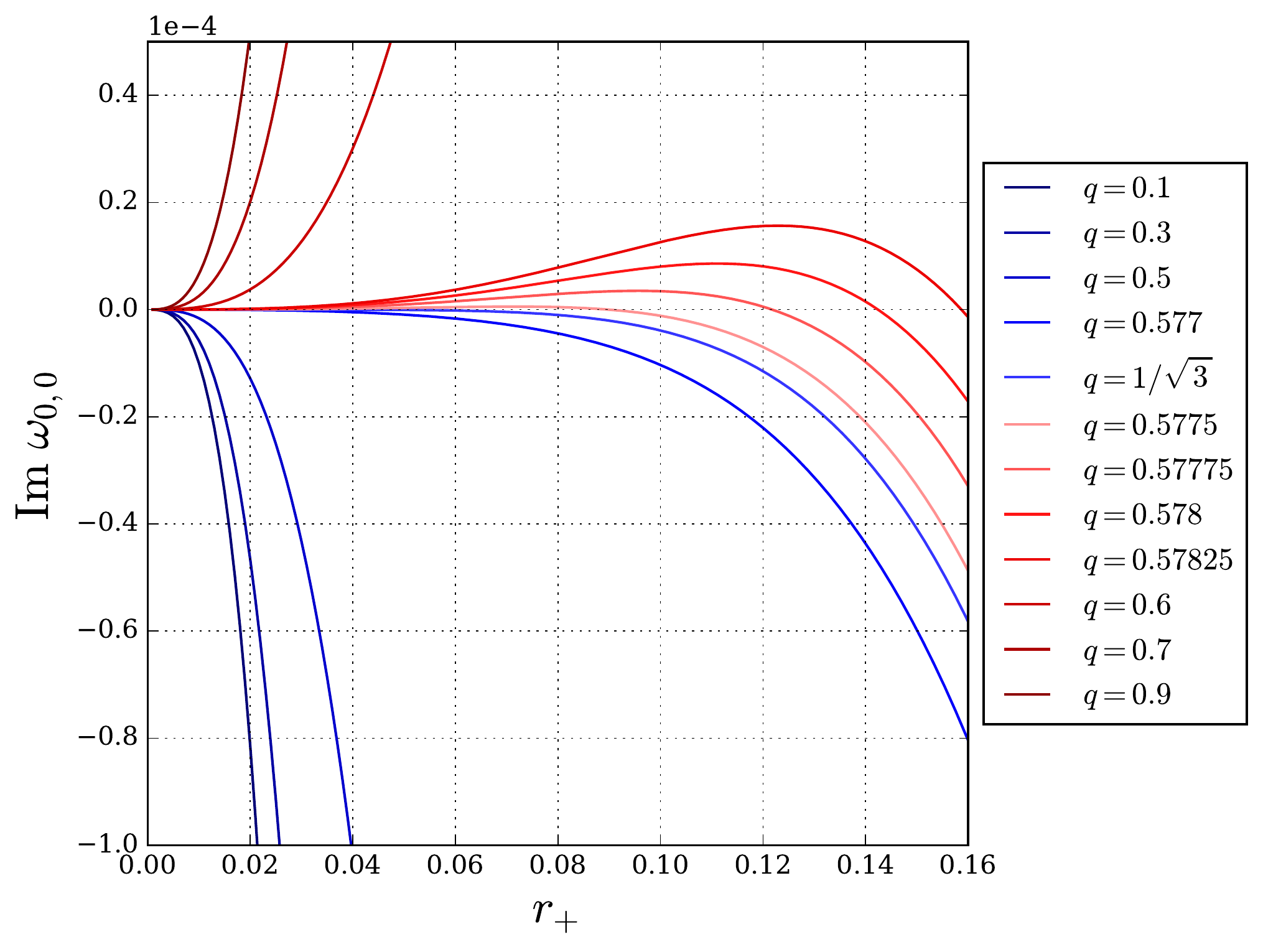}
\caption{Numerical results for the imaginary part of the fundamental QNM $\omega_{0,0}$ as a function of $r_{+}$ for increasing values of $q$ in the interval $\left[0.1,0.9\right]$. The charge and the scaling dimension of the scalar field are $e=8$ and $\Delta=4$, respectively.}\label{fig:im_omega_x_r2}
\end{figure}

Figure \ref{fig:im_omega_x_r2} shows our numerical solutions for the imaginary part of $\omega_{0,0}$ and illustrates explicitly that for $q=\{0.5775,0.57775,0.578,0.57825\}$ the superradiance condition is satisfied in a finite interval of $r_{+}$, and the stability is eventually restored as the black hole radius increases. As $q \rightarrow 1$, the superradiance condition is always satisfied in the small $r_{+}$ limit.
The QNMs frequencies shown in Fig. \ref{fig:im_omega_x_r2} have been obtained through the numerical implementation of the initial conditions for the PVI tau function using an arbitrary-precision library in \href{http://julialang.org}{Julia} programming language \cite{Painleve}. For the fast convergence of the numerical results, we have introduced the Fredholm determinant formulation of the PVI tau function, instead of its combinatorial description. We refer to previous works \cite{Amado:2017kao,Barragan-Amado:2018pxh,Anselmo20180080} for a more detailed discussion of the numerical methods.

Finally, in terms of $\epsilon = 1 - q$ and for $n=0$, \eqref{eq:omegan0} reads
\begin{equation}\label{eq:omega00_finiteT}
\omega_{0,0} = \Delta - \frac{1}{2}(\Delta-1)(\Delta(\epsilon^{2}-2\epsilon+2)-\sqrt{3}e(1-\epsilon))r_{+}^{2} - i\frac{1}{2}(\Delta-1)(2\Delta-\sqrt{3}e(1-\epsilon))r_{+}^{3} + \mathcal{O}(r_{+}^{4},r_{+}^{4}\log r_{+}),
\end{equation}
where $e < 2(\Delta+2)/\sqrt{3}$. Equation \eqref{eq:omega00_finiteT} should provide a good description of the high-temperature regime $r_{+} \ll \epsilon < 1$. We perform a detailed comparison of \eqref{eq:omega00_finiteT} with the numerical solution in Sec. \ref{sec:3c} (Fig. \ref{fig:analytical_and_numerical_epsilon}).

\subsection{The low-temperature limit of the fundamental QNM}
\label{sec:3b}
The description of the low-temperature regime requires a more careful approach since the single monodromy parameters \eqref{eq:smallr2_monodromies} may diverge a priori. This issue is solved by means of a double series expansion for small $r_{+}$ and $\epsilon$, and by considering the scaling regime $\epsilon \ll r_{+} \ll 1$, which indeed corresponds to the low-temperature regime of small black holes, i.e. near-extremal small black holes. In the low-temperature limit \eqref{eq:twopitemp} is given by
\begin{equation}\label{eq:low_temp}
2\pi T_{+} = \frac{2\epsilon}{r_{+}} + \mathcal{O}\left(\frac{\epsilon^{2}}{r_+},\epsilon r_{+}\right), 
\end{equation}
and the single monodromies in terms of $\epsilon$ and $r_{+}$ are
\begin{equation}\label{eq:thetas_lowtemp}
\begin{split}
&\theta_{+}=i\dfrac{(2\omega-\sqrt{3}e)}{4}\dfrac{r_{+}}{\epsilon}+i\dfrac{1}{16}(2\omega+\sqrt{3}e)(2+\epsilon)r_{+} + \mathcal{O}\left(\epsilon^{2} r_{+},\frac{r_{+}^{3}}{\epsilon},r_{+}^{3},\epsilon r_{+}^{3}\right), \\ 
&\theta_{-}= -i\dfrac{(2\omega-\sqrt{3}e)}{4}\dfrac{r_{+}}{\epsilon}+i\dfrac{1}{8}(10\omega-3\sqrt{3}e)r_{+} - i\dfrac{1}{16}(14\omega-\sqrt{3}e)\epsilon r_{+} + \mathcal{O}\left(\epsilon^{2}r_{+},\frac{r_{+}^{3}}{\epsilon},r_{+}^{3},\epsilon r_{+}^{3}\right), \\
&\theta_{0}=\omega-\dfrac{1}{2}(6\omega-\sqrt{3}e)(1-\epsilon)r_{+}^{2}+\mathcal{O}(\epsilon^{2}r_{+}^{2},r_{+}^{4},\epsilon r_{+}^{4}),
\end{split}
\end{equation}
where $\theta_{\pm}$ diverge as $\epsilon \rightarrow 0$, as it is also evident from the pole at $q^{2}=1$ in \eqref{eq:smallr2_monodromies}. 

The initial conditions for the confluent limit of the PVI tau function were derived in \cite{BarraganAmado:2021uyw}. The procedure is further analogous to the rotating case in \cite{BarraganAmado:2021uyw}. We need to compute the correction of order $r_{+}^{2}$, as well as the first non-trivial correction in $\epsilon$ for $\omega_{0,0} = \Delta + \eta_{0,0}r_{+}^{2}$ in \eqref{eq:omega} and 
\begin{equation}\label{eq:sigma00}
\sigma_{0,0} \equiv \sigma_{0}(\omega=\omega_{0,0}) = 2 - \nu_{0,0}r_{+}^{2}.
\end{equation}
For this purpose, we expand the left hand side of \eqref{eq:fourt0K0} in the small $\epsilon$ and $r_{+}$ limit to obtain
\begin{equation}\label{eq:lhsK0}
\begin{split}
4z_{0}K_{0} &= \frac{1}{4}i(6\Delta-\sqrt{3}e)(2-\epsilon)r_{+}+\frac{1}{8}i(2\Delta+\sqrt{3}e)\epsilon^{2}r_{+}+\frac{1}{16}(3e^{2}+4\sqrt{3}e\Delta+4\Delta(\Delta-16))(1-\epsilon)r_{+}^{2} \\
&+ \mathcal{O}(\epsilon^{3}r_{+},\epsilon^{2}r_{+}^{2},r_{+}^{3},\epsilon r_{+}^{3})
\end{split}
\end{equation}
while the right hand side of \eqref{eq:fourt0K0} gives 
\begin{equation}	\label{eq:rhsK0}
\begin{split}
4z_{0}K_{0} &= \frac{1}{4}i(6\Delta-\sqrt{3}e)(2-\epsilon)r_{+}+\frac{1}{8}i(2\Delta+\sqrt{3}e)\epsilon^{2}r_{+} \\
&- \biggl[ 2\nu_{0,0} -\frac{1}{16}(\sqrt{3}e-6\Delta)^{2} -4\nu_{0,0}x\left(1+x+2x^{2}+5x^{3}+\ldots\right)\biggr] r_{+}^{2} \\
&- \biggl[ \frac{1}{16}(\sqrt{3}e-6\Delta)^{2} + 8\nu_{0,0}x\left(1+2x+6x^{2}+20x^{3}+\ldots\right)\biggr]\epsilon r_{+}^{2} + \mathcal{O}(\epsilon^{3}r_{+},\epsilon^{2}r_{+}^{2},r_{+}^{3},\epsilon r_{+}^{3}),
\end{split}
\end{equation}
where the dots are the higher order poles in $\sigma_{0,0}$ that contribute to the $r_{+}^{2}$ and $\epsilon r_{+}^{2}$ terms, and we have defined
\begin{equation}
x = \frac{(2\Delta-\sqrt{3}e)(\sqrt{3}e-6\Delta)(\Delta-1)}{16\nu_{0,0}^{2}}.
\end{equation}
Equating \eqref{eq:lhsK0} and \eqref{eq:rhsK0} leads to the following equation for $\nu_{0,0}$
\begin{equation}
\frac{2\nu_{0,0}}{\sqrt{1-4x}} - \frac{8\nu_{0,0}x(1-\epsilon)}{\sqrt{1-4x}} = \Delta(2(\Delta+2)-\sqrt{3}e)(1-\epsilon) + \mathcal{O}(\epsilon^{2}),
\end{equation}
that can be solved to yield
\begin{equation}
\nu_{0,0} = \frac{1}{2}(1 - \epsilon)\sqrt{4\Delta^{2}(\Delta^{2}+\Delta+7)-4\sqrt{3}e\Delta(\Delta^{2}+2)+3e^{2}(\Delta^{2}-\Delta+1)} + \mathcal{O}(\epsilon^{2}).
\end{equation}
Hence, $\sigma_{0,0}$ in \eqref{eq:sigma00} reads
\begin{equation}\label{eq:sigma00}
\begin{split}
\sigma_{0,0} &= 2 - \nu_{0,0}r_{+}^{2} \\
&= 2 - \tilde{\nu}_{0,0}(1 - \epsilon)r_{+}^{2} + \mathcal{O}(\epsilon^{2}r_{+}^{2}),
\end{split}
\end{equation} 
where
\begin{equation}\label{eq:nu00}
\tilde{\nu}_{0,0} = \frac{1}{2}\sqrt{4\Delta^{2}(\Delta^{2}+\Delta+7)-4\sqrt{3}e\Delta(\Delta^{2}+2)+3e^{2}(\Delta^{2}-\Delta+1)}.
\end{equation}
The asymptotic expansion for $\chi_{0,0} \equiv \chi_{0}(\omega=\omega_{0,0})$ now receives contributions from all orders in $z_{0}$ due to the pole structure at $\sigma=2$ in \eqref{eq:kappaexpansion}. 
Nevertheless, one can again use the generating function for the Catalan numbers to obtain an analytic expansion for $\epsilon \ll r_{+} \ll 1$ given by 
\begin{equation}\label{eq:chi00}
\chi_{0,0} = \frac{1}{2} + \frac{(\Delta-1)(2\Delta-\sqrt{3}e)(6\Delta-\sqrt{3}e)}{16\tilde{\nu}_{0,0}^{2}}
+\frac{1}{2}\sqrt{1+\frac{(\Delta-1)(2\Delta-\sqrt{3}e)(6\Delta-\sqrt{3}e)}{4\tilde{\nu}_{0,0}^{2}}} + \mathcal{O}(\epsilon^{2},r_{+}^{2},\epsilon r_{+}^{2}).
\end{equation}
By means of the expansion for $\theta_{0}$ in \eqref{eq:thetas_lowtemp}, with $\omega=\omega_{0,0}$ in \eqref{eq:omega}, we obtain
\begin{equation}
\begin{split}
\theta_{0} &= \Delta - \beta_{0,0}r_{+}^{2}\\
&= \Delta - \left(\frac{1}{2}(6\Delta-\sqrt{3}e)(1-\epsilon) - \eta_{0,0}\right)r_{+}^{2},
\end{split}
\end{equation}
where
\begin{equation}\label{eq:eta00}
\eta_{0,0} = \frac{1}{2}(1-\epsilon)(6\Delta-\sqrt{3}e) - \beta_{0,0}.
\end{equation}
Inserting \eqref{eq:chi00} in the right hand side of \eqref{eq:painleve} for $n=\ell=0$ and expanding the left hand side of \eqref{eq:painleve} for $\epsilon \ll r_{+} \ll 1$, we obtain
\begin{equation}\label{eq:for_beta}
\begin{split}
&\frac{(\Delta-1)(6\Delta-\sqrt{3}e)(2\Delta-\sqrt{3}e)}{16\tilde{\nu}_{0,0}^{2}}\frac{\beta_{0,0}+\tilde{\nu}_{0,0}}{\beta_{0,0}-\tilde{\nu}_{0,0}}\biggl[1 -\frac{2\tilde{\nu}_{0,0}\beta_{0,0}\epsilon}{\beta_{0,0}^{2}-\tilde{\nu}_{0,0}^{2}} + \frac{i4\tilde{\nu}_{0,0}r_{+}}{(6\Delta-\sqrt{3}e)}
-\frac{i2\tilde{\nu}_{0,0}(\beta_{0,0}^{2}+4\beta_{0,0}\tilde{\nu}_{0,0}-\tilde{\nu}_{0,0}^{2})\epsilon r_{+}}{(6\Delta-\sqrt{3}e)(\beta_{0,0}^{2}-\tilde{\nu}_{0,0}^{2})}\\ 
&+ \mathcal{O}(\epsilon^{2},\epsilon^{2}r_{+},r_{+}^{2},r_{+}^{2}\log r_{+},\epsilon r_{+}^{2},\epsilon r_{+}^{2}\log r_{+})\biggr]\\
&= \frac{1}{2} + \frac{(\Delta-1)(2\Delta-\sqrt{3}e)(6\Delta-\sqrt{3}e)}{16\tilde{\nu}_{0,0}^{2}}
+\frac{1}{2}\sqrt{1+\frac{(\Delta-1)(2\Delta-\sqrt{3}e)(6\Delta-\sqrt{3}e)}{4\tilde{\nu}_{0,0}^{2}}}
+ \mathcal{O}(\epsilon^{2},r_{+}^{2},\epsilon r_{+}^{2})
\end{split}
\end{equation}
whose perturbative solution for $\beta_{0,0}$ reads
\begin{equation}\label{eq:beta00}
\beta_{0,0} = \frac{1}{2}\Delta(2(\Delta+2)-\sqrt{3}e)(1-\epsilon) + i\frac{1}{4}(\Delta-1)(2\Delta-\sqrt{3}e)(2-3\epsilon)r_{+} + \mathcal{O}(\epsilon^{2},\epsilon^{2}r_{+},r_{+}^{2},r_{+}^{2}\log r_{+},\epsilon r_{+}^{2},\epsilon r_{+}^{2}\log r_{+}). 
\end{equation}
Therefore, by means of \eqref{eq:eta00} and \eqref{eq:beta00}, the fundamental QNM frequency is
\begin{equation}\label{eq:omega00}
\begin{gathered}
\omega_{0,0} = \Delta - \frac{1}{2}(\Delta-1)(2\Delta-\sqrt{3}e)(1-\epsilon)r_{+}^{2} -i\frac{1}{4}(\Delta-1)(2\Delta-\sqrt{3}e)(2-3\epsilon)r_{+}^{3} + \mathcal{O}(\epsilon^{2}r_{+}^{2},r_{+}^{4},r_{+}^{4}\log r_{+},\epsilon r_{+}^{4},\epsilon r_{+}^{4}\log r_{+}),
\end{gathered}
\end{equation}
which is valid for small black holes in the low-temperature limit, i.e. $\epsilon \ll r_{+} \ll 1$. The imaginary part of $\omega_{0,0}$ in \eqref{eq:omega00}, at this order of perturbation theory, suggests that the black hole is stable $(\mathrm{Im}\,\omega_{0,0} < 0)$ under perturbations with charge $e < 2\Delta/\sqrt{3}$ and $\Delta > 1$. The imaginary part has a leading correction in $\epsilon$ of order $\epsilon r_{+}^{3}$, indicating that contributions from the black hole temperature are not sufficiently enhanced to trigger superradiant instabilities for $e < 2\Delta/\sqrt{3}$. Such instabilities can instead be triggered by the charge of the scalar field provided that $\Delta > 1$. Incidentally, we notice that although equations \eqref{eq:omegan0} and \eqref{eq:omega00} were derived in two different regimes, they coincide in the extremal limit $\epsilon = 0$ $(q=1)$ at least at the given order in $r_{+}$.

\subsection{Comparison with numerical results}
\label{sec:3c}
It is interesting to compare the analytical expression \eqref{eq:omega00} for small black holes at low temperature with the numerical prediction for a given charge of the scalar field $e$ as a function of the radius of the outer horizon $r_{+}$, as well as the $\epsilon$ parameter. We recall that the numerical results have been obtained by solving the transcendental equations \eqref{eq:initial_value} for the Fredholm determinant of the PVI tau function \cite{Gavrylenko:2016zlf}. The initial conditions have been implemented in \href{http://julialang.org}{Julia}, using an arbitrary-precision library. Then, the QNM frequencies have been found by applying a root-finding algorithm which employs the Newton-Raphson method.

\begin{figure}
\centering
\includegraphics[width=\textwidth]{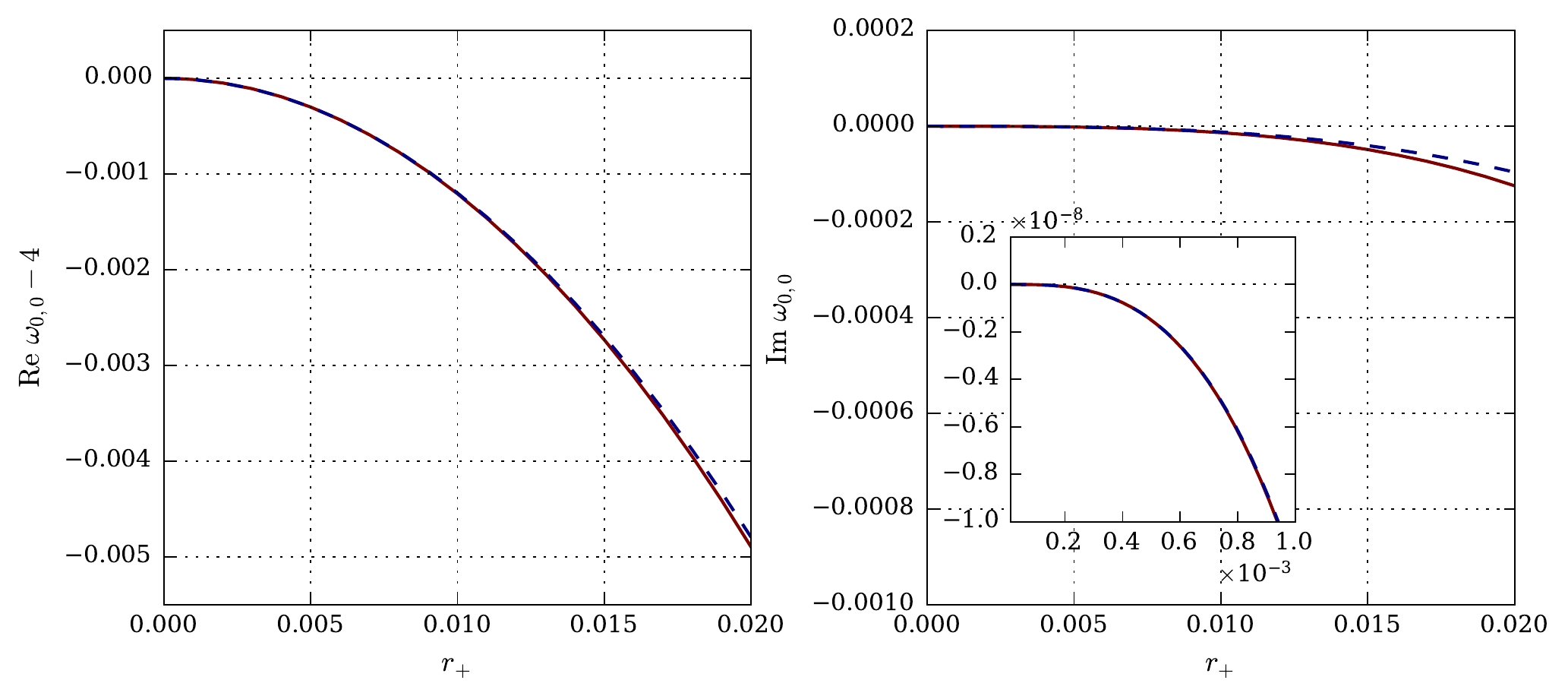}
\caption{The analytical result (blue dashed line) in \eqref{eq:omega00} for the real (left) and imaginary (right) part of the fundamental QNM frequency $\omega_{0,0}$ compared with the numerical result (red solid line) as function of $r_{+}$ for fixed $\epsilon=5 \times 10^{-6}$ and charge of the scalar field $e = 0.005$.}
\label{fig:analytical_and_numerical_rplus}
\end{figure}

\begin{figure}
\centering
\includegraphics[width=\textwidth]{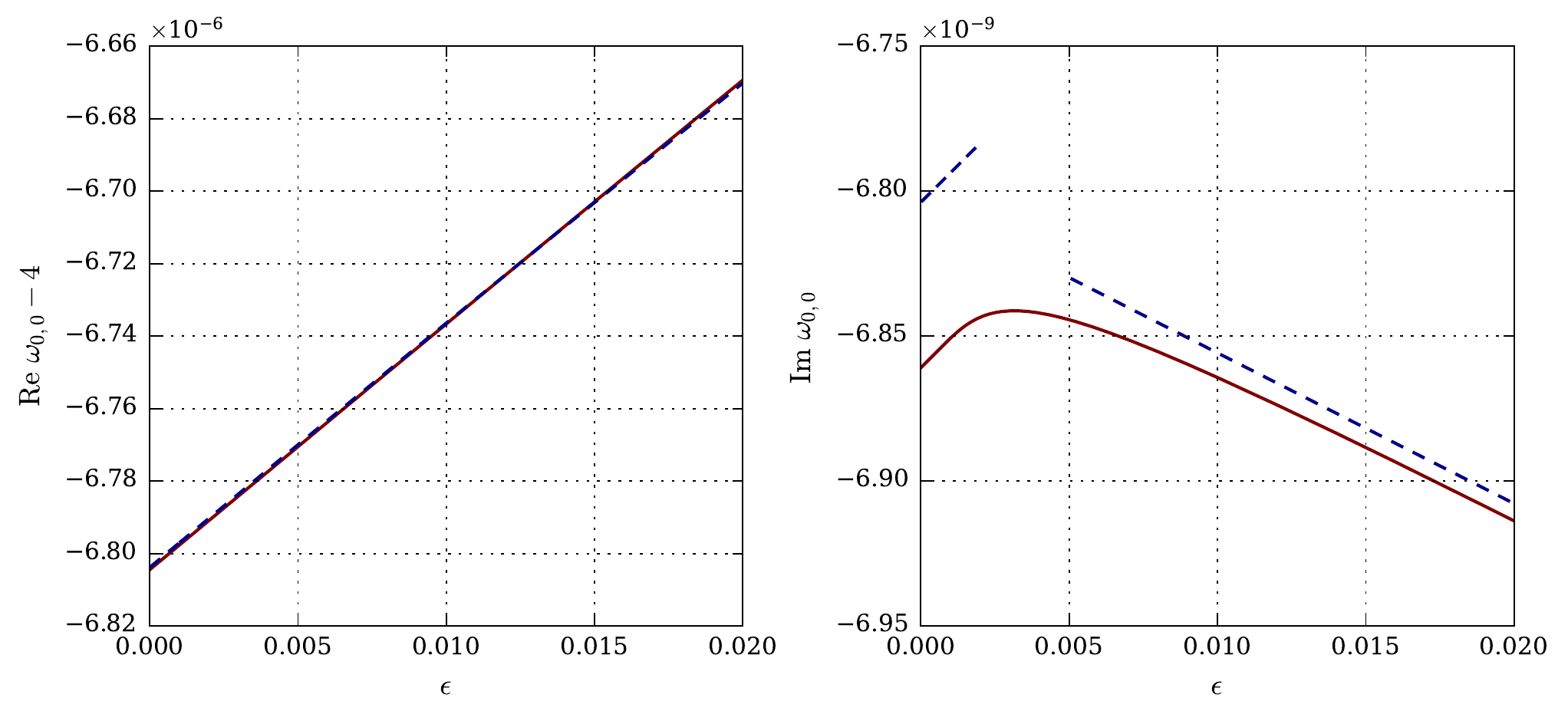}
\caption{The analytical result (blue dashed line) in \eqref{eq:omega00} for the real (left) and imaginary (right) part of the fundamental QNM frequency $\omega_{0,0}$ compared with the numerical result (red solid line) as function of $\epsilon$ for $e=2$ and $r_{+}=0.001$. The ascending and descending dashed lines for the $\mathrm{Im}\,\omega_{0,0}$ in the right panel reproduce \eqref{eq:omega00} and \eqref{eq:omega00_finiteT}, respectively. The ascending line describes the same slope of the numerical result, however, the shift w.r.t. the numerical result is due to higher orders in $r_+$, for $\epsilon=0$ in \eqref{eq:omega00}.}
\label{fig:analytical_and_numerical_epsilon}
\end{figure}

Figure \ref{fig:analytical_and_numerical_rplus} shows the comparison between the numerical results (in red, solid line) and the low-temperature asymptotic formula \eqref{eq:omega00} for the fundamental QNM frequency as a function of $r_{+}$ for fixed small $\epsilon = 5\times 10^{-6}$ and the charge of the scalar field $e = 0.005$. We observe a small discrepancy between the two curves, both for the real and imaginary part of $\omega_{0,0}$, increasing to about $2\,\%$ for $r_{+} \simeq 0.01$. In Figure \ref{fig:analytical_and_numerical_epsilon} we display the transition from low to high temperatures — small and large $\epsilon$, respectively — by varying $\epsilon$ while keeping $r_{+}=0.001$ fixed. The numerical results (again in red, solid line) are compared with the asymptotic formulas (in blue, dashed line) for high and low temperature, given by \eqref{eq:omega00_finiteT} and \eqref{eq:omega00}, respectively.  For low-temperatures, we have $\epsilon \ll r_{+} \ll 1$ and thus $r_{+}^{3}$ corrections to \eqref{eq:omega00} are more relevant for the approximation than the $\epsilon r_+^2$ terms. In other words, one should resum the series in $r_{+}$ at each given order in $\epsilon$ for a full comparison. 
Nevertheless, the $\epsilon$ corrections encode the temperature dependence, and can be checked to agree with the numerical prediction by comparing the slope of the two curves for small $\epsilon$ in the imaginary part of $\omega_{0,0}$ in Fig. \ref{fig:analytical_and_numerical_epsilon}.

\section{Discussion}
\label{sec:4}

In this paper, we have investigated the quasinormal modes of a charged scalar field scattering on small RN-AdS$_{5}$ black holes, where the outer horizon radius $r_{+}$ is much smaller than the AdS radius, via the isomonodromy method. We analytically explored two regimes in terms of $r_{+}$ and the parameter $\epsilon = 1 - q$, with $0 \leq q \leq 1$ the charge of the black hole in units of the critical charge $Q_{c}$. In section \ref{sec:3a}, we have considered the small black hole approximation at high temperatures $(r_{+} \ll \epsilon < 1)$ and provided an asymptotic expression for the QNMs frequencies for $n$ generic and $\ell=0$ that generalizes previous results found in the literature \cite{Wang:2014eha}. 
These asymptotic expressions display instabilities for the fundamental QNM mode for large enough $q$, and we verified numerically (for $e=8$ and $\Delta=4$) that there is a superradiance window whose width in $r_+$ increases with $q$. 
We leave open the question whether, in parallel to the Hawking-Page transition for uncharged black holes \cite{Hawking1982}, stability might be restored as the radius of the black hole increases. 

A different approach is required in the low-temperature limit $\epsilon \ll r_{+} \ll 1$ and we have presented the analysis of this case in section \ref{sec:3b}. By considering a double series expansion for small $\epsilon$ and $r_{+}$, we have computed the first non-trivial correction in $\epsilon$ to the fundamental QNM frequency. We found that small black holes at low temperature are stable under scalar perturbations with $\Delta$ satisfying the unitarity bound $\Delta > 1$, and charge $e < 2\Delta/\sqrt{3}$. When the latter conditions are not met, the imaginary part of \eqref{eq:omega00} becomes positive, meaning that the black hole is unstable.
   
In \cite{Wang:2014eha}, the correction to the empty AdS frequency is derived from the matching condition between the analytical solutions constructed in terms of hypergeometric functions. In contrast, we have shown via the isomonodromy method that the asymptotic expansion for the fundamental QNM frequency receives contributions from all descendants of the CFT primaries in the vacuum module. These contributions can be resummed in terms of the generating functions of the Catalan numbers and the result suggests that in the matching approach one should consider all levels in the Frobenius expansion. It is worth mentioning that for numerical purposes the Fredholm determinant representation of the PVI tau function can reach a faster convergence time than direct instanton counting methods based on Nekrasov functions. 

This work can be seen as an application to RN-AdS$_{5}$ of the ideas exposed in \cite{BarraganAmado:2021uyw}, where the authors have written the initial conditions for the Painlev\'{e} V tau function to compute an asymptotic expansion for the fundamental QNM frequency in the low-temperature regime.

Finally, in \cite{PhysRevD.81.124020} it was argued that large RN-AdS$_{4}$ black holes, i.e $r_{+} \gg 1$ in units of the AdS curvature radius, become unstable below a critical temperature. Since one can control $\epsilon$ and $r_{+}$ in the accessory parameter expansion, it would be interesting to explore this regime and the nature of these instabilities in terms of the PVI tau function and its confluent limits. It is also interesting to study the hydrodynamical modes in RN-AdS$_{5}$ black holes, since scalar perturbations in this background have been introduced as dual CFT states of strongly correlated holographic models with finite density, for which the bound on the speed of sound can be violated \cite{PhysRevD.94.106008}. The method presented here and in \cite{BarraganAmado:2021uyw} works for higher spin perturbations on rotating black holes \cite{Amado:2020zsr} and can be used to study those perturbations in RN-AdS$_{5}$. 

\section*{Acknowledgments}
J.B.A. is grateful for the support from the University of Sherbrooke as well as from the Mathematical Physics Laboratory at the Centre de Recherches Mathématiques in Montréal. 

\appendix

\section{Nekrasov expansion for the PVI tau function}
\label{appendixA}

The series representation of the PVI tau function, written in \cite{Gamayun:2012ma,Gamayun:2013auu}, around the critical point $t=0$ is given by
\begin{equation}
\tau(t)=\sum_{n\in\mathbb{Z}}C(\vec{\theta},\sigma+2n)s^nt^{\tfrac{1}{4}((\sigma+2n)^2-\theta_1^2-\theta_2^2)}\mathcal{B}(\vec{\theta},\sigma+2n;t),
\label{eq:nekrasovexpansion}
\end{equation}
where $\vec{\theta}=\lbrace\theta_1,\theta_2,\theta_3,\theta_{4}\rbrace$ are the single monodromy parameters, and the parameters $\sigma$, $s$ are two integration constants. The structure constants $C(\vec{\theta},\sigma)$ are given in terms of Barnes' functions
\begin{equation}
C(\vec{\theta},\sigma)=\frac{\prod_{\alpha,\beta=\pm}
	G(1+\tfrac{1}{2}( \theta_{3}+\alpha\theta_{4}+\beta\sigma))
	G(1+\tfrac{1}{2}( \theta_{2}+\alpha\theta_{1}+\beta\sigma))}{G(1+\sigma)G(1-\sigma)},
\end{equation}
and $\mathcal{B}(\vec{\theta},\sigma;t)$ in \eqref{eq:nekrasovexpansion} coincides with the $c=1$ Virasoro conformal blocks, which are explicitly given by a combinatorial series in terms of Nekrasov functions
\begin{equation}\label{eq:CB}
\mathcal{B}(\vec{\theta},\sigma+2n;t) = (1-t)^{\tfrac{1}{2}\theta_{2}\theta_{3}}\sum_{\lambda,\mu\in \mathbb{Y}}{\cal B}_{\lambda,\mu}(\vec{\theta},\sigma+2n) t^{|\lambda|+|\mu|},
\end{equation}
where $\mathbb{Y}$ denotes the space of Young diagrams, $\lambda$ and $\mu$ are two of its elements, with number of boxes $|\lambda|$ and $|\mu|$, and the coefficients in \eqref{eq:CB} are
\begin{equation}
\begin{split}
  {\cal B}_{\lambda,\mu}(\vec{\theta},\sigma) &=
  \prod_{(i,j)\in\lambda}\frac{((\theta_t+\sigma+2(i-j))^2-\theta_0^2)
    ((\theta_1+\sigma+2(i-j))^2-\theta_\infty^2)}{16h_\lambda^2(i,j)
    (\lambda'_j-i+\mu_i-j+1+\sigma)^2} \\
  &\times \prod_{(i,j)\in\mu}\frac{((\theta_t-\sigma+2(i-j))^2-\theta_0^2)
    ((\theta_1-\sigma+2(i-j))^2-\theta_\infty^2)}{16h_\lambda^2(i,j)
    (\mu'_j-i+\lambda_i-j+1-\sigma)^2}.
\end{split}
\end{equation}
For each box situated at $(i,j)$ in $\lambda$, $\lambda_i$ is the number of boxes at row $i$ of $\lambda$ and $\lambda'_j$ is the number of boxes at column $j$ of $\lambda$; $h(i,j)=\lambda_i+\lambda'_j-i-j+1$ is the hook length of the box at $(i,j)$. The parameter $s$ can be determined in terms of the monodromy matrices $\lbrace \sigma, \sigma_{23} \rbrace$ from the formula $(3.48a)$ in \cite{ILP2016}:
\begin{align}
\sin^2\pi\sigma\cos\pi\sigma_{23} &=
\cos\pi\theta_1\cos\pi\theta_4+
\cos\pi\theta_2\cos\pi\theta_3 \nonumber\\
&-\cos\pi\sigma(\cos\pi\theta_1\cos\pi\theta_3+
\cos\pi\theta_2\cos\pi\theta_4) \nonumber\\
&-\frac{1}{2}(\cos\pi\theta_4-\cos\pi(\theta_3-\sigma))
(\cos\pi\theta_1-\cos\pi(\theta_2-\sigma))s \nonumber\\  
&-\frac{1}{2}(\cos\pi\theta_4-\cos\pi(\theta_3+\sigma))
(\cos\pi\theta_1-\cos\pi(\theta_2+\sigma))
s^{-1}.  
\label{eq:sigma1t}
\end{align}
For $t$ sufficiently close to zero\footnote{One can choose different series expansions around the other critical points of the PVI tau function at $t=1$ and $t=\infty$ \cite{Jimbo:1982}.}, and generic monodromy parameters in the sense that
\begin{equation}\label{eq:monoconds}
\sigma \notin \mathbb{Z}, \qquad \sigma \pm \theta_{1} \pm \theta_{2} \notin \mathbb{Z}, \qquad \sigma \pm \theta_{3} \pm \theta_{4} \notin \mathbb{Z},
\end{equation}
we have
\begin{align}\label{eq:asymp_tau}
\tau(t) = C_{0}\,&t^{\tfrac{1}{4}(\sigma^{2}-\theta^{2}_{1}-\theta^{2}_{2})}(1-t)^{\tfrac{1}{2}\theta_{3}\theta_{2}}\biggl\{1+\bigg[\frac{\theta_{3}\theta_{2}}{2}+\frac{(\theta^{2}_{1}-\theta^{2}_{2}-\sigma^{2})(\theta^{2}_{4}-\theta^{2}_{3}-\sigma^{2})}{8\sigma^{2}}\nonumber \\
&-\frac{(\theta^{2}_{1}-(\theta_{2}-\sigma)^{2})(\theta^{2}_{4}-(\theta_{3}-\sigma)^{2})}{16\sigma^{2}(1+\sigma)^{2}}\kappa\,t^{\sigma}-\frac{(\theta^{2}_{1}-(\theta_{2}+\sigma)^{2})(\theta^{2}_{4}-(\theta_{3}+\sigma)^{2})}{16\sigma^{2}(1-\sigma)^{2}}\frac{1}{\kappa\,t^{\sigma}}\bigg]t + \cdots\biggr\},
\end{align}
where $0 < \mathrm{Re}\,\sigma < 1$, $C_{0}$ is a constant independent of $t$, and $\kappa$ is a known function of the monodromy parameters:
\begin{equation}
  \kappa=s\frac{\Gamma^2(1-\sigma)}{\Gamma^2(1+\sigma)}
  \frac{\Gamma(1+\tfrac{1}{2}(\theta_2+\theta_1+\sigma))
    \Gamma(1+\tfrac{1}{2}(\theta_2-\theta_1+\sigma))}{
    \Gamma(1+\tfrac{1}{2}(\theta_2+\theta_1-\sigma))
    \Gamma(1+\tfrac{1}{2}(\theta_2-\theta_1-\sigma))}
    \frac{\Gamma(1+\tfrac{1}{2}(\theta_3+\theta_4+\sigma))
    \Gamma(1+\tfrac{1}{2}(\theta_3-\theta_4+\sigma))}{
    \Gamma(1+\tfrac{1}{2}(\theta_3+\theta_4-\sigma))
    \Gamma(1+\tfrac{1}{2}(\theta_3-\theta_4-\sigma))}.
\label{eq:kappa}
\end{equation}

\bibliography{Biblio_RNAdS}

\end{document}